\documentclass[
preprint,
showpacs,
showkeys
]{revtex4-1}
\usepackage{amsthm}
\theoremstyle{plain}
\newtheorem*{theorem*}{Theorem}

\newtheorem*{definition*}{Definition}

\newtheorem*{lemma*}{Lemma}

\usepackage{braket}
\usepackage{delarray}
\usepackage{here}

\usepackage{txfonts}

\usepackage[dvipdfmx]{graphicx}
\usepackage{bm}
\usepackage{color}
\usepackage{ulem}

\newcommand{\be}{\begin{eqnarray}}
\newcommand{\ee}{\end{eqnarray}}
\newcommand{\ba}{\begin{array}}
\newcommand{\ea}{\end{array}}
\newcommand{\bmat}{\left(\begin{array}}
\newcommand{\emat}{\end{array}\right)}

\begin{document}
\title{
Probability bounds on trajectory-level exceedances of\\ quantum speed limits  under continuous-measurement feedback
}

\author{Kohei Kobayashi}

\affiliation{
Department of Physics, Institute of Science Tokyo, 2-12-1 Ookayama, Meguro-ku, Tokyo, Japan
}

\begin{abstract} 
We investigate the probability that a conditioned quantum
trajectory exceeds an ensemble-derived quantum speed limit
(QSL) velocity under continuous measurement-based feedback.
Using the finite-time average fidelity velocity,
we derive an upper bound on the exceedance probability
in terms of its mean and variance.
We further obtain state-independent bounds from uniform
constraints on the fidelity drift and diffusion,
including an exponential concentration bound.
For a feedback-controlled qubit, we derive explicit
model-specific bounds and show that they are nontrivial
in a short-time regime.
We also identify a finite-time regime in which
the exceedance probability vanishes exactly due to
the restricted support of the conditioned dynamics.
Monte Carlo simulations confirm the analytical bounds
and illustrate their tightness.


\end{abstract}
\date{\today}
\maketitle

\section{Introduction}

Quantum speed limits (QSLs) provide fundamental bounds on
the rate of quantum-state evolution.
Since the seminal works of Mandelstam and Tamm~\cite{MT}
and Margolus and Levitin~\cite{ML},
QSLs have been extensively studied in various settings,
including open quantum systems [3-7],
driven dynamics [8-11],
quantum metrology~\cite{Giovanetti,Beau},
quantum computation~\cite{Lloyd},
and quantum control [15-18].
In particular, QSLs provide useful benchmarks for
evaluating the performance of quantum control protocols.

Continuous measurement-based feedback (MBF) control
uses measurement records to adjust the control input
in real time [19-25].
The dynamics of such systems can be described at two levels:
the ensemble-averaged state and the conditioned quantum
trajectories associated with individual measurement records.
While open-system QSLs constrain the ensemble-averaged
evolution, conditioned trajectories fluctuate due to
measurement backaction.
Consequently, individual trajectories may exhibit
finite-time average velocities exceeding a QSL velocity
derived from the ensemble dynamics.
Such exceedances do not constitute violations of a QSL
for the conditioned dynamics, but reflect the distinction
between ensemble and trajectory descriptions.

Trajectory-level exceedances of ensemble-derived QSL
velocities were investigated numerically in
Ref.~\cite{Pintos}, where fluctuations in trajectory
velocities and passage times were characterized.
However, analytical bounds on the probability of
such exceedances were not established.
Moreover, the analysis did not include
measurement-record-dependent feedback.
In MBF systems, the control input can depend on the
conditioned state, introducing correlations between
the feedback and measurement noise.
These correlations make the trajectory statistics
nontrivial and generally prevent the ensemble dynamics
from forming a closed master equation.

In this paper, we derive analytical upper bounds on
the probability that a conditioned trajectory exceeds
an ensemble-derived QSL velocity under continuous MBF.
We consider the finite-time average decrease of the
initial-state fidelity, which defines a random velocity
over a fixed observation interval.
Applying Cantelli's inequality, we obtain a general
probability bound determined by the mean and variance
of this velocity.
We further derive state-independent bounds using
uniform constraints on the fidelity drift and diffusion.
These include both a variance-based bound and an
exponential concentration bound, which can be evaluated
without determining the full trajectory distribution
once the ensemble QSL velocity is known.

We apply these results to a continuously monitored
qubit stabilization model.
Exploiting the geometry of the conditioned dynamics,
we obtain explicit model-specific bounds and identify
a short-time regime in which the state-independent
concentration bounds are nontrivial.
Furthermore, the restricted support of the conditioned
state yields a finite-time regime in which the
exceedance probability vanishes exactly.
We compare these analytical predictions with
Monte Carlo simulations.

The remainder of this paper is organized as follows.
In Sec.~II, we derive the general probability bounds
for continuously monitored feedback dynamics.
In Sec.~III, we apply the results to a feedback-controlled
qubit and examine their analytical and numerical
properties.

\section{Probability bounds for continuously monitored quantum trajectories}
\label{sec:main_result}

\subsection{Stochastic fidelity dynamics}

We consider a quantum system subject to continuous measurement and
measurement-based feedback (MBF). Its conditioned state $\rho_t^{\rm c}$
obeys the stochastic master equation (SME)
\begin{equation}
 d\rho_t^{\rm c}=-iu_t[H,\rho_t^{\rm c}]\,dt
 +\mathcal D[L]\rho_t^{\rm c}\,dt
 +\mathcal H[L]\rho_t^{\rm c}\,dW_t,
 \label{sme}
\end{equation}
where $H$ is a Hermitian control Hamiltonian, $u_t$ is a real-valued
feedback input adapted to the measurement record, and $L=L^\dagger$ is
the measurement operator. We set $\hbar=1$ and use
\begin{equation}
 \mathcal D[L]\rho=-\frac12[L,[L,\rho]],\qquad
 \mathcal H[L]\rho=\{L,\rho\}-2{\rm Tr}(L\rho)\rho.
\end{equation}
The innovation process $dW_t=dy_t-2{\rm Tr}(L\rho_t^{\rm c})\,dt$
is a Wiener increment satisfying $\mathrm E[dW_t]=0$ and $(dW_t)^2=dt$.
Here $\mathrm E[\cdots]$ denotes averaging over measurement records.

The ensemble state $\rho_t=\mathrm E[\rho_t^{\rm c}]$ evolves as
\begin{equation}
 \dot\rho_t=-i[H,\mathrm E(u_t\rho_t^{\rm c})]
 +\mathcal D[L]\rho_t.
 \label{ensemble_dynamics}
\end{equation}
This equation need not be closed, since the feedback and the conditioned
state can be correlated.

For a pure initial state $\rho_0=|\psi_0\rangle\langle\psi_0|$, we define
the conditioned initial-state fidelity
\begin{equation}
 F_t^{\rm c}={\rm Tr}(\rho_0\rho_t^{\rm c}),
 \qquad F_0^{\rm c}=1.
 \label{conditioned_fidelity}
\end{equation}
Equation~(\ref{sme}) gives
\begin{equation}
 dF_t^{\rm c}=A(\rho_t^{\rm c})\,dt
 +B(\rho_t^{\rm c})\,dW_t,
 \label{fidelity_sde}
\end{equation}
with
\begin{eqnarray}
 A(\rho)&=&-iu_t{\rm Tr}([\rho_0,H]\rho)
 -\frac12{\rm Tr}([L,[L,\rho_0]]\rho),
 \label{fidelity_drift}\\
 B(\rho)&=&{\rm Tr}(\{\rho_0,L\}\rho)
 -2{\rm Tr}(\rho_0\rho){\rm Tr}(L\rho).
 \label{fidelity_diffusion}
\end{eqnarray}
The drift $A$ may itself depend on the measurement record through $u_t$.
The ensemble fidelity is
\begin{equation}
 F_t=\mathrm E[F_t^{\rm c}]={\rm Tr}(\rho_0\rho_t).
 \label{ensemble_fidelity}
\end{equation}

\subsection{Finite-time exceedance probability}

To compare single trajectories with the ensemble evolution, define the
finite-time average fidelity velocity over $[0,\tau]$ by
\begin{equation}
 v_{\rm c}=\frac{1-F_\tau^{\rm c}}{\tau}
 =\overline A+\overline B,
 \label{finite_time_average_velocity}
\end{equation}
where
\begin{eqnarray}
 \overline A&=&-\frac1\tau\int_0^\tau
 A(\rho_s^{\rm c})\,ds,
 \label{finite_time_average_drift}\\
 \overline B&=&-\frac1\tau\int_0^\tau
 B(\rho_s^{\rm c})\,dW_s.
 \label{finite_time_stochastic_contribution}
\end{eqnarray}
Since the It\^o integral has zero mean,
$\mathrm E[v_{\rm c}]=\mathrm E[\overline A]$
and
\begin{equation}
 {\rm Var}(v_{\rm c})
 ={\rm Var}(\overline A)
 +\frac1{\tau^2}\mathrm E\!\left[\int_0^\tau
 B^2(\rho_s^{\rm c})\,ds\right]
 +2{\rm Cov}(\overline A,\overline B).
 \label{finite_time_variance_decomposition}
\end{equation}
In general, the covariance term does not vanish under feedback.

For the ensemble trajectory, Schatten-norm duality yields
\begin{equation}
 F_0-F_\tau
 =-\int_0^\tau{\rm Tr}(\rho_0\dot\rho_s)\,ds
 \leq\int_0^\tau\|\dot\rho_s\|_\infty\,ds.
 \label{ensemble_fidelity_speed_bound}
\end{equation}
We therefore use the finite-time ensemble QSL reference velocity
\begin{equation}
 v_{\rm QSL}=\frac1\tau\int_0^\tau
 \|\dot\rho_s\|_\infty\,ds,
 \label{qsl_velocity}
\end{equation}
which satisfies
\begin{equation}
 \mathrm E[v_{\rm c}]=\frac{F_0-F_\tau}{\tau}
 \leq v_{\rm QSL}.
 \label{mean_velocity_qsl}
\end{equation}
Here $\|\cdot\|_\infty$ denotes the operator norm. The event
$v_{\rm c}\geq v_{\rm QSL}$ is an exceedance of an ensemble-derived
reference by a conditioned trajectory, not a violation of a QSL for the
conditioned dynamics.

By applying Cantelli's inequality, i.e., the one-sided Chebyshev inequality,
\cite{Cantelli,Boucheron} to $v_{\rm c}$, we obtain
\begin{equation}
 \Pr[v_{\rm c}\geq v_{\rm QSL}]
 \leq\frac{{\rm Var}(v_{\rm c})}
 {{\rm Var}(v_{\rm c})+
 [v_{\rm QSL}-\mathrm E(v_{\rm c})]^2},
 \label{main_bound_finite_time}
\end{equation}
provided $v_{\rm QSL}>\mathrm E[v_{\rm c}]$ and the variance is finite.
If the ensemble evolution saturates the reference bound, Cantelli's
inequality yields only the trivial upper bound $1$.

\subsection{State-independent concentration bounds}

The bound~(\ref{main_bound_finite_time}) requires trajectory statistics.
We now derive bounds that depend instead on uniform drift and diffusion
estimates. Assume, almost surely for $0\leq t\leq\tau$, that
\begin{equation}
 -A(\rho_t^{\rm c})\leq A_{\max},\qquad
 B^2(\rho_t^{\rm c})\leq B_{\max}^2,
 \label{state_independent_conditions}
\end{equation}
with $B_{\max}>0$. Set $\delta=v_{\rm QSL}-A_{\max}$.
When $\delta>0$, the event $v_{\rm c}\geq v_{\rm QSL}$ implies
$\overline B\geq\delta$. By It\^o isometry, $ {\rm E}[\overline B]=0$ and we find
\begin{equation}
 {\rm Var}(\overline B)
 =\frac1{\tau^2}\mathrm E\!\left[\int_0^\tau
 B^2(\rho_s^{\rm c})\,ds\right]
 \leq\frac{B_{\max}^2}{\tau}.
 \label{state_independent_noise_variance}
\end{equation}
Cantelli's inequality therefore gives
\begin{equation}
 \Pr[v_{\rm c}\geq v_{\rm QSL}]
 \leq\frac{B_{\max}^2}{B_{\max}^2+\tau\delta^2}.
 \label{Cantelli bound}
\end{equation}

An exponential estimate follows from the exponential martingale
associated with $-\int_0^\tau B(\rho_s^{\rm c})\,dW_s$.
For every $\lambda>0$, the exponential Markov inequality and
Eq.~(\ref{state_independent_conditions}) imply
\begin{equation}
 \Pr[\overline B\geq\delta]
 \leq\exp\!\left(-\lambda\tau\delta
 +\frac{\lambda^2B_{\max}^2\tau}{2}\right).
\end{equation}
Choosing $\lambda=\delta/B_{\max}^2$ yields
\begin{equation}
 \Pr[v_{\rm c}\geq v_{\rm QSL}]
 \leq\exp\!\left(-\frac{\tau\delta^2}{2B_{\max}^2}\right).
 \label{state_independent_exponential_bound}
\end{equation}
Combining the two estimates, we obtain
\begin{equation}
\Pr[v_{\rm c}\geq v_{\rm QSL}]
 \leq\min\!\left\{
 \frac{B_{\max}^2}{B_{\max}^2+\tau\delta^2},
 \exp\!\left(-\frac{\tau\delta^2}{2B_{\max}^2}\right)
 \right\},
 \label{combined_state_independent_bound}
\end{equation}
where $\delta=v_{\rm QSL}-A_{\max}>0$.
The first estimate is tighter near $\delta=0$, while the exponential
estimate dominates sufficiently far into the tail.

For bounded feedback $|u_t|\leq u_{\max}$, explicit choices of the
constants are obtained from the spectral widths
$r_X=\lambda_{\max}(X)-\lambda_{\min}(X)$ ($X=H,L$):
\begin{eqnarray}
B_{\max}^2&=&\frac{r_L^2}{4}, \\
 A_{\max}&=&\frac{u_{\max}r_H}{2}
 +\frac{r_L}{2}\Delta_{|\psi_0\rangle}L,
 \label{spectral_drift_diffusion_bounds}
\end{eqnarray}
where
$\Delta_{|\psi_0\rangle}L=
\sqrt{\langle\psi_0|L^2|\psi_0\rangle
 -\langle\psi_0|L|\psi_0\rangle^2}$.
The derivation is given in Appendix~\ref{app:state_independent_derivation}.
Substitution into Eq.~(\ref{combined_state_independent_bound}) gives
\begin{equation}
 \Pr[v_{\rm c}\geq v_{\rm QSL}]
 \leq\min\{P_{\rm C},P_{\rm E}\},
 \label{spectral_probability_bound}
\end{equation}
where
\begin{eqnarray}
 P_{\rm C}&=&\frac{r_L^2}
 {r_L^2+\tau(2v_{\rm QSL}-u_{\max}r_H
 -r_L\Delta_{|\psi_0\rangle}L)^2},\\
 P_{\rm E}&=&\exp\!\left[
 -\frac{\tau(2v_{\rm QSL}-u_{\max}r_H
 -r_L\Delta_{|\psi_0\rangle}L)^2}{2r_L^2}
 \right].
 \label{state_independent_bound_rL}
\end{eqnarray}
These expressions apply when $v_{\rm QSL}>A_{\max}$ (and $r_L>0$).
They require neither the distribution nor the moments of the
conditioned velocity once $v_{\rm QSL}$ is known; for feedback control,
the ensemble reference $v_{\rm QSL}$ may still need to be evaluated
from the dynamics. The quantum-specific ingredients are the fidelity
process, the ensemble QSL reference, and the operator bounds, rather
than the concentration inequalities themselves.

\section{Example: two-level feedback stabilization}

\subsection{Qubit feedback model and fidelity dynamics}

We apply the finite-time probability bounds to a qubit
feedback-stabilization model.
The two-level system consists of the states
$|0\rangle=(1,0)^\top$ and $|1\rangle=(0,1)^\top$.
We consider
\begin{eqnarray}
H=\sigma_y,\ \ \ L
=
\sqrt{\kappa}\sigma_z,
\label{qubit_operators}
\end{eqnarray}
where $\{\sigma_x,\sigma_y,\sigma_z\}$ are the Pauli matrices.
This setting is commonly used in quantum feedback control
\cite{Handel,Vijay}.
The operator $L$ describes the continuous measurement of $\sigma_z$, with
measurement strength $\kappa$, while the feedback control is applied through
the Hamiltonian $u_tH$.

We choose the feedback input as
\begin{eqnarray}
u_t=-\alpha\kappa x_t^{\rm c},
\label{feedback_law_qubit}
\end{eqnarray}
where $\alpha>0$ is a dimensionless feedback gain and
$x_t^{\rm c}={\rm Tr}
\left(
\sigma_x\rho_t^{\rm c}
\right)$,
$y_t^{\rm c}={\rm Tr}
\left(
\sigma_y\rho_t^{\rm c}
\right)$,
and $z_t^{\rm c}=
{\rm Tr}
\left(
\sigma_z\rho_t^{\rm c}
\right)$.
The feedback protocol is designed to stabilize the target state $|0\rangle$.
We choose the initial state as
$|\psi_0\rangle
=(|0\rangle+|1\rangle)/\sqrt{2}$ and
$\rho_0=|\psi_0\rangle\langle\psi_0|=
(I+\sigma_x)/2$.
For $\alpha>2$, the target state is asymptotically stabilized almost surely (the proof is given in Appendix B1).

Although the feedback protocol stabilizes $|0\rangle$, the fidelity used in
the QSL analysis is defined with respect to the initial state.
It is given by
\begin{eqnarray}
F_t^{\rm c}
=
{\rm Tr}
\left(
\rho_0\rho_t^{\rm c}
\right)
=\frac{1+x_t^{\rm c}}{2}.
\label{qubit_initial_fidelity}
\end{eqnarray}

As shown in Appendix~\ref{app:qubit_model_derivation}, the stochastic
dynamics of $F_t^{\rm c}$ is governed by
\begin{eqnarray}
dF_t^{\rm c}
=
-\kappa x_t^{\rm c}
\left(
1+\alpha z_t^{\rm c}
\right)dt
-\sqrt{\kappa}
x_t^{\rm c}z_t^{\rm c}dW_t.
\label{qubit_fidelity_sde}
\end{eqnarray}
Comparison with Eq.~(\ref{fidelity_sde}) gives
\begin{eqnarray}
A(\rho_t^{\rm c})&=&-\kappa x_t^{\rm c}
\left(
1+\alpha z_t^{\rm c}
\right),  \\ 
B(\rho_t^{\rm c})
&=&
-\sqrt{\kappa}
x_t^{\rm c}z_t^{\rm c}.
\end{eqnarray}

Substituting this into the general formula in
Eq. (28) gives the
state-dependent probability bound for the present model. 
Its direct
evaluation requires the finite-time statistics of the conditioned dynamics.
Therefore, we focus below on model-specific state-independent bounds.

We next derive model-specific state-independent bounds.
For unit-efficiency monitoring, a pure initial state remains pure.
Moreover, the plane $y_t^{\rm c}=0$ is invariant, and the present initial
condition satisfies $y_0^{\rm c}=0$.
Consequently,
$(x_t^{\rm c})^2+(z_t^{\rm c})^2=1$.
The solution also satisfies $x_t^{\rm c}>0$ at every finite time almost
surely.
These properties are shown in
Appendix~\ref{app:qubit_model_derivation}.

The diffusion coefficient satisfies
\begin{eqnarray}
B^2(\rho_t^{\rm c})
=\kappa
(x_t^{\rm c})^2
(z_t^{\rm c})^2
\leq
\frac{\kappa}{4},
\end{eqnarray}
and hence we may choose
$B_{\max}^2=\kappa/4$.

The instantaneous drift of the departure velocity satisfies
\begin{eqnarray}
-A(\rho_t^{\rm c})
=\kappa x_t^{\rm c}
\left(
1+\alpha z_t^{\rm c}
\right)
\leq
\kappa M_\alpha,
\end{eqnarray}
where
$M_\alpha =
\sqrt{1-z_*^2}
\left(
1+\alpha z_*
\right)$ and
$z_*=\frac{
\sqrt{1+8\alpha^2}-1
}{
4\alpha
}$.
Therefore, a valid model-specific drift bound is
$A_{\max}=\kappa M_\alpha$.
The derivation of the model-specific bounds
$B_{\max}^2=\kappa/4$ and $A_{\max}=\kappa M_\alpha$
is given in Appendix B 2.

For the present model, $P_{\mathrm C}$ and $P_{\mathrm E}$ are
\begin{eqnarray}
P_{\mathrm C}
&=&
\frac{
\kappa
}{
\kappa
+
4\tau
\left(
v_{\rm QSL}
-\kappa M_\alpha
\right)^2}, \\
P_{\mathrm E}
&=&
\exp
\left[
-\frac{
2\tau
\left(
v_{\rm QSL}
-\kappa M_\alpha
\right)^2
}{
\kappa
}
\right].
\end{eqnarray}
Combining the two estimates, we obtain
\begin{eqnarray}
{\rm Pr}
\left[
v_{\rm c}
\geq
v_{\rm QSL}
\right]
\leq
\min
\left\{
P_{\mathrm C},
P_{\mathrm E}
\right\},
\label{qubit_state_independent_bound}
\end{eqnarray}
provided that
$v_{\rm QSL}>\kappa M_\alpha$.

The expressions give a closed analytical
upper bound once the reference velocity $v_{\rm QSL}$ has been determined
from the ensemble dynamics.
The quantity $\kappa M_\alpha$ bounds the deterministic rate at which a
trajectory departs from the initial state, while $\kappa/4$ bounds the
measurement-induced fluctuation strength.

These model-specific estimates are tighter than the general spectral
estimates because they use the purity of the conditioned state and the
invariant geometry of the Bloch dynamics.
For fixed $\kappa$, $\alpha$, and $v_{\rm QSL}$, both
$P_{\mathrm C}$ and $P_{\mathrm E}$ decrease with the observation time
$\tau$.
However, changing $\kappa$ modifies both the maximum deterministic drift
and the maximum stochastic fluctuation strength.
Therefore, the dependence of the resulting probability bound on the measurement
strength is not generally monotonic.

\subsection{Analytical regimes of the model-specific probability bounds}

To characterize the parameter regimes in which the model-specific bounds
are nontrivial, we introduce the dimensionless quantities
\begin{eqnarray}
s=\kappa\tau,\ \ 
q=\frac{v_{\rm QSL}}{\kappa}.
\label{dimensionless_parameters}
\end{eqnarray}
In terms of these variables,
$P_{\rm C}$ and $P_{\rm E}$ become
\begin{eqnarray}
P_{\mathrm C}
&=&
\frac{1}{
1+4s\left(q-M_\alpha\right)^2
},
\\
P_{\mathrm E}
&=&
\exp
\left[
-2s\left(q-M_\alpha\right)^2
\right],
\end{eqnarray}
provided that
$q>M_\alpha$.

The relative strength of the two bounds can be characterized analytically.
Defining
$\chi=4s\left(q-M_\alpha\right)^2$,
we have
$P_{\mathrm C}
=(1+\chi)^{-1}$, and
$P_{\mathrm E}=
e^{-\chi/2}$.
The two bounds coincide at $\chi=0$.
Their nonzero crossing point is
$\chi_{\rm cr}\simeq2.513$.
Therefore, $P_{\mathrm C}$ is tighter for
$0<\chi<\chi_{\rm cr}$, whereas $P_{\mathrm E}$ is tighter for
$\chi>\chi_{\rm cr}$.
Thus, the exponential estimate is particularly useful when the QSL threshold
lies sufficiently far above the maximum deterministic departure rate.

A stronger model-specific restriction follows from the finite-time support
of the initial-state fidelity.
As shown in Appendix~\ref{app:qubit_model_derivation},
$x_t^{\rm c}>0$ at every finite time almost surely.
Hence, Eq.~(\ref{qubit_initial_fidelity}) gives
$F_\tau^{\rm c}>1/2$ at every finite $\tau$ almost surely.
Consequently,
\begin{eqnarray}
v_{\rm c}
=\frac{1-F_\tau^{\rm c}}{\tau}
<
\frac{1}{2\tau}
\label{qubit_velocity_support}
\end{eqnarray}
almost surely.
It follows that
${\rm Pr}
\left[
v_{\rm c}\geq v_{\rm QSL}
\right]
=0$
if $q\geq1/(2s)$.
Combining this support restriction with the concentration bounds, 
we obtain
the following fully analytical upper bound:
\begin{equation}
{\rm Pr}
\left[
v_{\rm c}\geq v_{\rm QSL}
\right]
\leq
\left\{
\begin{array}{ll}
0,
&
q\geq {1\over 2s}, \\[1.0ex]
\min\left\{
P_{\rm C},
P_{\rm E}
\right\},
&
M_\alpha<q<{1\over 2s}, \\[1.0ex]
1,
&
q\leq M_\alpha \ {\rm and}\ q<{1\over 2s}.
\end{array}
\right.
\end{equation}
The intermediate regime in which the state-independent concentration bounds
are nontrivial exists only when
$s<1/(2M_\alpha)$.

This zero-probability regime is not a consequence of a generic concentration
inequality. It follows from the finite-time support of the conditioned
quantum trajectory. More concretely, the pure conditioned state remains on
the unit circle in the $x$-$z$ plane of the Bloch sphere, and the solution
satisfies $x_t^c>0$ at every finite time. These state-space constraints imply
$F_\tau^c>1/2$ and hence impose the exact cutoff $v_c<1/(2\tau)$.
In this sense, the model-specific content is not limited to the numerical
values of $A_{\max}$ and $B_{\max}$.

We next show that the condition $q>M_\alpha$ is satisfied by the actual
ensemble-level QSL velocity in the short-time regime.
Averaging the Bloch equations derived in
Appendix~\ref{app:qubit_model_derivation} gives
\begin{eqnarray}
\frac{d{\rm E}[x^c_t]}{dt}
&=&
-2\kappa
{\rm E}
\left[
x_t^{\rm c}
\left(
1+\alpha z_t^{\rm c}
\right)
\right],
\\
\frac{d{\rm E}[z^c_t]}{dt}
&=&
2\alpha\kappa
{\rm E}
\left[
(x_t^{\rm c})^2
\right].
\end{eqnarray}
Since the ensemble-averaged state remains in the $x$-$z$ plane,
$\rho_t=
\left(
I+ {\rm E}[x^c_t]\sigma_x
+{\rm E}[z^c_t]\sigma_z
\right)/2$,
and hence
\begin{eqnarray}
\left\|
\dot{\rho}_t
\right\|_\infty
=\kappa
\sqrt{
{\rm E}
\left[
x_t^{\rm c}
\left(
1+\alpha z_t^{\rm c}
\right)
\right]^2
+
\alpha^2
{\rm E}
\left[
(x_t^{\rm c})^2
\right]^2
}.
\label{qubit_ensemble_speed}
\end{eqnarray}

At the initial time,
$x_0^{\rm c}=1$ and $z_0^{\rm c}=0$, so that
\begin{eqnarray}
\frac{d{\rm E}[x^c_t]}{dt} \biggr|_{t=0}
=-2\kappa, \ \ 
\frac{d{\rm E}[z^c_t]}{dt}\biggr|_{t=0}
=
2\alpha\kappa.
\end{eqnarray}
Therefore,
\begin{equation}
\left\|
\left.
\frac{d\rho_t}{dt}
\right|_{t=0}
\right\|_\infty
=
\kappa\sqrt{1+\alpha^2}.
\end{equation}
By continuity of the ensemble dynamics,
\begin{eqnarray}
\lim_{s\rightarrow0^+}
q(s)=\sqrt{1+\alpha^2}.
\label{short_time_q_limit}
\end{eqnarray}

On the other hand,
$M_\alpha$ is the maximum of
$\sqrt{1-z^2}(1+\alpha z)$ over $z\in[-1,1]$.
At the maximizing point,
the Cauchy--Schwarz inequality gives
\begin{eqnarray}
M_\alpha
&=&
\sqrt{1-z_*^2}
\left(
1+\alpha z_*
\right) \nonumber \\
&\leq&
\sqrt{1+\alpha^2}
\sqrt{
1-z_*^4
}
\nonumber\\
&<&
\sqrt{1+\alpha^2},
\label{Malpha_short_time_comparison}
\end{eqnarray}
where the final inequality holds for $\alpha>0$.
Equations~(\ref{short_time_q_limit})
and~(\ref{Malpha_short_time_comparison}) imply that
$q>M_\alpha$
for sufficiently small $s$.
Thus, the condition required for the state-independent concentration bounds
is not merely formal; it is satisfied by the actual QSL velocity in a
finite short-time regime.

Finally, the actual value of $q$ is itself bounded by the conditioned-state
geometry.
For $\alpha>0$ and $z\in[-1,1]$,
\begin{eqnarray}
\left|
\sqrt{1-z^2}
\left(
1+\alpha z
\right)
\right|
\leq
M_\alpha.
\end{eqnarray}
Together with
$(x_t^{\rm c})^2\leq1$,
Eq.~(\ref{qubit_ensemble_speed}) therefore gives
\begin{eqnarray}
\left\|
\dot{\rho}_t
\right\|_\infty
\leq
\kappa
\sqrt{
M_\alpha^2+\alpha^2
}.
\end{eqnarray}
Because $v_{\rm QSL}:=(1/\tau)\int^\tau_0\|\dot{\rho}_s\|_{\infty}ds$, 
we obtain
\begin{equation}
q\leq\sqrt{
M_\alpha^2+\alpha^2}.
\end{equation}
This result shows that $q$ is constrained by the actual ensemble dynamics
and should not be regarded as an arbitrary independent parameter.
The three regimes in Eq.~(64) are illustrated in Fig. 1.
The figure shows the trivial-bound region, the nontrivial concentration-bound
region, and the zero-probability region in the dimensionless $(s,q)$ plane.

\begin{figure}[t]
    \centering
    \includegraphics[width=0.8\textwidth]{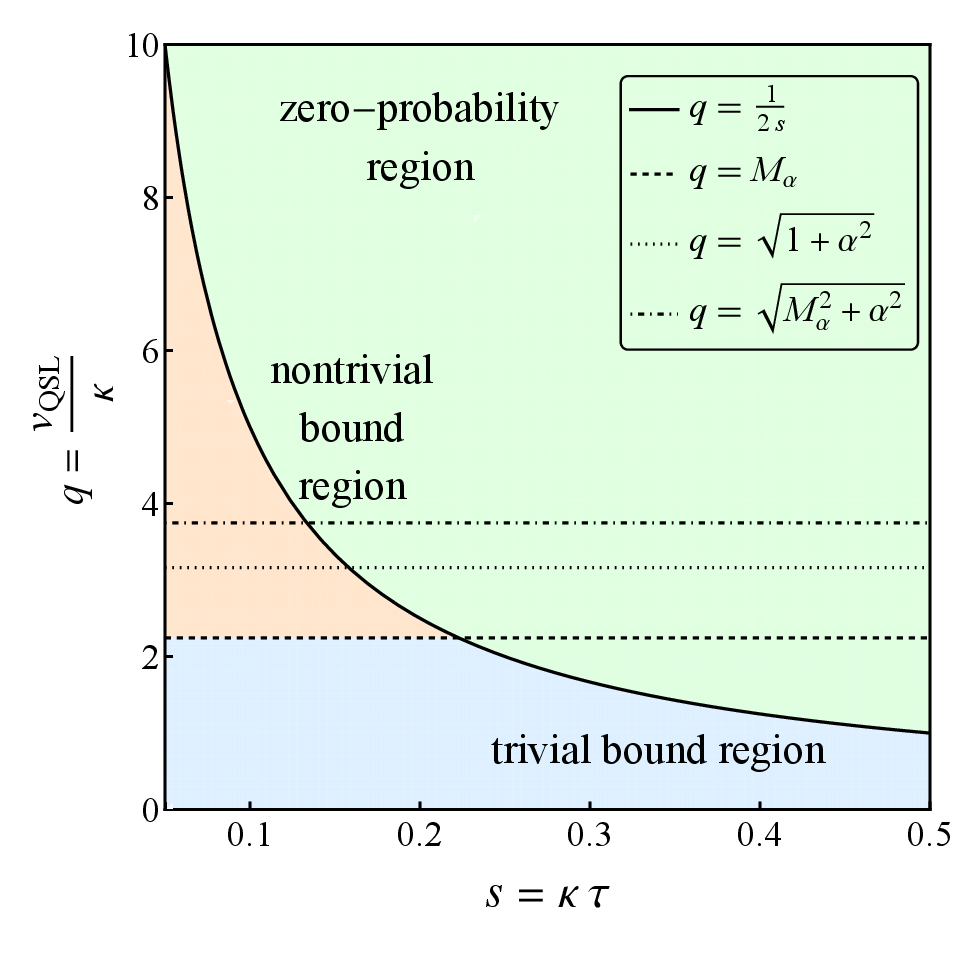}
    \caption{
    Analytical regimes of the model-specific probability bound in the
    $(s,q)$ plane for $\alpha=3$.
    The solid line $q=1/(2s)$ represents the finite-support cutoff, above
    which the exceedance probability vanishes exactly.
    The dashed line $q=M_\alpha$ gives the threshold above which the
    state-independent concentration bounds become nontrivial.
    The dotted and dash-dotted lines show the short-time
    limit $q\to\sqrt{1+\alpha^2}$ and the analytical upper bound
    $q\le\sqrt{M_\alpha^2+\alpha^2}$ for the ensemble-level QSL velocity, respectively.
    The lower blue, middle orange, and upper green regions correspond to the
trivial-bound, nontrivial-bound, and zero-probability regimes,
respectively. }
    \label{fig:sq-regime}
\end{figure}

To examine the quantitative performance of the obtained bounds, we also
performed Monte Carlo simulations of the conditioned qubit dynamics.
For numerical stability, we integrated the dimensionless logit equation
in Eq.~(B4), with the rescaled time $s=\kappa t$, by the Euler-Maruyama
method. 
We used $5.0\times10^4$ trajectories and a
dimensionless time step $10^{-3}$. 
The dimensionless reference
velocity $q(s)=v_{\mathrm{QSL}}/\kappa$ was estimated from the ensemble
of simulated trajectories, and the exceedance probability was evaluated
as the fraction of trajectories satisfying $v_{\rm c}\ge v_{\mathrm{QSL}}$.

\begin{figure}[t]
\centering
\includegraphics[width=0.8\textwidth]{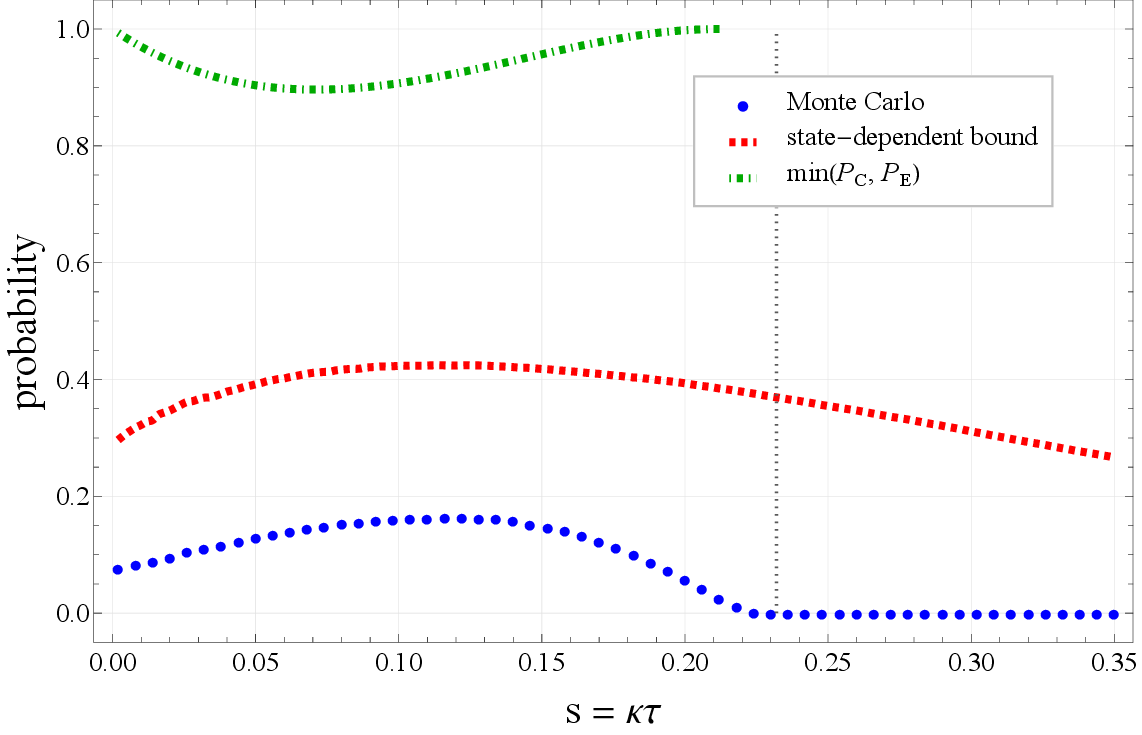}
\caption{
Monte Carlo comparison of the empirical exceedance probability with the
probability bounds for the qubit feedback model at $\alpha=3$.
The simulation uses $5.0\times10^4$ trajectories and a dimensionless time
step $10^{-3}$.
Blue symbols represent the probability
${\rm Pr}[v_{\rm c} \ge v_{\mathrm{QSL}}]$ obtained from the simulated
trajectories. The red dashed line shows the state-dependent bound,
evaluated using the Monte Carlo estimates of the finite-time mean
and variance of $v_{\rm c}$. The green dash-dotted line shows the
state-independent bound $\min\{P_{\rm C},P_{\rm E}\}$.
The vertical dotted line marks the finite-support cutoff.
}
\label{fig:simulation}
\end{figure}

\section{Conclusion}

We have derived analytical upper bounds on the probability
that a conditioned quantum trajectory exceeds an
ensemble-derived QSL velocity under continuous
measurement-based feedback.
For the finite-time average fidelity velocity,
we obtained a general bound in terms of its mean
and variance, as well as state-independent bounds
based on uniform constraints on the drift and diffusion.
The latter include an exponential concentration bound
and can be evaluated without determining the full
trajectory distribution once the ensemble QSL velocity
is known.

Applying these results to a feedback-controlled qubit,
we derived explicit model-specific bounds and identified
a short-time regime in which the state-independent
bounds are nontrivial.
We also established a finite-time regime in which
the exceedance probability vanishes exactly due to
the restricted support of the conditioned dynamics.
Monte Carlo simulations support these analytical results.
Our findings provide a probabilistic characterization
of trajectory-level deviations from ensemble-derived
QSL benchmarks without implying any violation of
quantum speed limits for conditioned dynamics.

This work was supported by JSPS KAKENHI Grant No. 25H01360.

\appendix

\section{State-independent drift and diffusion bounds}
\label{app:state_independent_derivation}
Here we derive the state-independent bounds.
The diffusion coefficient of the initial-state fidelity is
\begin{eqnarray}
B(\rho)
=
{\rm Tr}
\left(
\{\rho_0,L\}\rho
\right)
-
2{\rm Tr}(\rho_0\rho)
{\rm Tr}(L\rho).
\label{appendix_diffusion_coefficient}
\end{eqnarray}
For a fixed state $\rho$, we introduce the centered operators
\begin{eqnarray}
\widetilde{\rho}_0 &=&
\rho_0
-
\langle\rho_0\rangle_\rho I, \\
\widetilde{L}
&=&
L-\langle L\rangle_\rho I,
\end{eqnarray}
where
$\langle O\rangle_\rho={\rm Tr}(O\rho)$.
Equation~(\ref{appendix_diffusion_coefficient}) 
can then be written as
\begin{eqnarray}
B(\rho)
=
\left\langle
\{
\widetilde{\rho}_0,
\widetilde{L}
\}
\right\rangle_\rho.
\end{eqnarray}

Using the Cauchy--Schwarz inequality for the symmetric covariance, we obtain
\begin{eqnarray}
B^2(\rho)
\leq
4
\left(
\Delta_\rho\rho_0
\right)^2
\left(
\Delta_\rho L
\right)^2,
\label{appendix_covariance_bound}
\end{eqnarray}
where
$\left(
\Delta_\rho O
\right)^2
=
\left\langle
\left(
O-\langle O\rangle_\rho I
\right)^2
\right\rangle_\rho$.

Since the initial state is pure,
$\rho_0=|\psi_0\rangle\langle\psi_0|$,
it is a rank-one projector and satisfies
$\rho_0^2=\rho_0$.
Therefore,
\begin{eqnarray}
\left(
\Delta_\rho\rho_0
\right)^2
=
\langle\rho_0\rangle_\rho
-
\langle\rho_0\rangle_\rho^2
\leq
\frac{1}{4}.
\label{appendix_initial_projector_variance}
\end{eqnarray}

For the Hermitian operator $L$, let
$r_L=\lambda_{\max}(L)-\lambda_{\min}(L)$
be its spectral width.
For every density operator $\rho$,
\begin{eqnarray}
\left(
\Delta_\rho L
\right)^2
\leq
\frac{r_L^2}{4}.
\label{appendix_L_variance}
\end{eqnarray}
Substituting
(\ref{appendix_initial_projector_variance}) and
(\ref{appendix_L_variance}) into
Eq.~(\ref{appendix_covariance_bound}), we find
\begin{eqnarray}
B^2(\rho)
\leq
\frac{r_L^2}{4}.
\end{eqnarray}
Thus, a valid state-independent choice is
\begin{eqnarray}
B_{\max}^2
=
\frac{r_L^2}{4}.
\end{eqnarray}

The stochastic fidelity equation is
\begin{eqnarray}
dF_t^{\rm c}
=
A(\rho_t^{\rm c})\,dt
+
B(\rho_t^{\rm c})\,dW_t,
\end{eqnarray}
with
\begin{eqnarray}
A(\rho)
=
-iu_t
\left\langle
[\rho_0,H]
\right\rangle_\rho
-
\frac{1}{2}
\left\langle
[L,[L,\rho_0]]
\right\rangle_\rho.
\label{appendix_fidelity_drift}
\end{eqnarray}
Since the finite-time velocity measures the decrease of the initial-state
fidelity, the relevant drift is
\begin{eqnarray}
-A(\rho)
=
iu_t
\left\langle
[\rho_0,H]
\right\rangle_\rho
+
\frac{1}{2}
\left\langle
[L,[L,\rho_0]]
\right\rangle_\rho.
\label{appendix_departure_drift}
\end{eqnarray}

We assume that
$|u_t|\leq u_{\max}$.
The Robertson uncertainty relation gives
\begin{eqnarray}
\left|
iu_t
\left\langle
[\rho_0,H]
\right\rangle_\rho
\right|
\leq
2u_{\max}
\Delta_\rho\rho_0
\Delta_\rho H.
\end{eqnarray}
Since
$\Delta_\rho\rho_0
\leq1/2$ and
$\Delta_\rho H
\leq
r_H/2$,
where
$r_H=\lambda_{\max}(H)-\lambda_{\min}(H)$,
we obtain
\begin{eqnarray}
\left|
iu_t
\left\langle
[\rho_0,H]
\right\rangle_\rho
\right|
\leq
\frac{u_{\max}r_H}{2}.
\label{appendix_H_drift_bound}
\end{eqnarray}

We next consider the measurement contribution.
Define the Hermitian operator
$Y=-i[L,\rho_0]$.
Since
$[L,[L,\rho_0]]
=i[L,Y]$,
the Robertson uncertainty relation gives
\begin{eqnarray}
\left|
\frac{1}{2}
\left\langle
[L,[L,\rho_0]]
\right\rangle_\rho
\right|
&=&
\frac{1}{2}
\left|
\left\langle
[L,Y]
\right\rangle_\rho
\right| \nonumber \\
&\leq&
\Delta_\rho L
\Delta_\rho Y.
\end{eqnarray}
Using
$\Delta_\rho L
\leq r_L/2$ and 
$\Delta_\rho Y
\leq\|Y\|_\infty$,
we obtain
\begin{eqnarray}
\left|
\frac{1}{2}
\left\langle
[L,[L,\rho_0]]
\right\rangle_\rho
\right|
\leq
\frac{r_L}{2}
\left\|
-i[L,\rho_0]
\right\|_\infty.
\label{appendix_L_drift_bound}
\end{eqnarray}

For the pure initial state
$\rho_0=|\psi_0\rangle\langle\psi_0|$,
the operator norm satisfies
\begin{eqnarray}
\left\|
-i[L,\rho_0]
\right\|_\infty
&=&
\sqrt{
\langle\psi_0|L^2|\psi_0\rangle
-
\langle\psi_0|L|\psi_0\rangle^2
}
\nonumber\\
&=&
\Delta_{|\psi_0\rangle}L.
\end{eqnarray}
Therefore, we find
\begin{eqnarray}
-A(\rho)
\leq
\frac{u_{\max}r_H}{2}
+
\frac{r_L}{2}
\Delta_{|\psi_0\rangle}L.
\end{eqnarray}
Therefore, a valid state-independent drift bound is
\begin{eqnarray}
A_{\max}
=
\frac{u_{\max}r_H}{2}
+
\frac{r_L}{2}
\Delta_{|\psi_0\rangle}L.
\end{eqnarray}

If the control input $u_t$ is specified externally and does not depend on
the conditioned state or the measurement record, the departure drift can be
written as
\begin{eqnarray}
-A(\rho)
=
{\rm Tr}
\left(
G_t\rho
\right),
\end{eqnarray}
where
$G_t=iu_t[\rho_0,H]+
[L,[L,\rho_0]]/2$.
The operator $G_t$ is Hermitian.
Therefore,
\begin{eqnarray}
-A(\rho)
\leq
\lambda_{\max}(G_t).
\end{eqnarray}
A sharper constant bound over the interval $[0,\tau]$ is then
\begin{eqnarray}
A_{\max}
=
\sup_{0\leq t\leq\tau}
\lambda_{\max}(G_t).
\end{eqnarray}

For measurement-based feedback, however, $u_t$ generally depends on the
conditioned state or on the previous measurement record.
In that case, $A_{\rm max}$ 
remains valid whenever the feedback amplitude is uniformly bounded.

\section{Details of the qubit feedback model}
\label{app:qubit_details}

\subsection{Convergence to the target state}
\label{app:qubit_stabilization}

We first show that the feedback protocol asymptotically stabilizes the target
state $|0\rangle$.
To distinguish it from the initial-state fidelity used in the QSL analysis, we denote the fidelity with respect to the target state by
\begin{eqnarray}
F^c_{{\rm tar}, t}
=\langle 0|\rho_t^{\rm c}|0\rangle
=\frac{1+z^c_t}{2}.
\label{appendix_target_fidelity}
\end{eqnarray}

As shown in the next subsection, the conditioned trajectory remains pure and
in the $x$-$z$ plane.
Therefore,
$(x^c_t)^2+(z^c_t)^2=1$.
Using the feedback law
$u_t=-\alpha\kappa x^c_t$,
the target fidelity satisfies
\begin{eqnarray}
dF^c_{{\rm tar}, t}
&=&
4\alpha\kappa
F^c_{{\rm tar}, t}
\left(
1-F^c_{{\rm tar}, t}
\right)dt +
4\sqrt{\kappa}
F^c_{{\rm tar}, t}
\left(
1-F^c_{{\rm tar}, t}
\right)dW_t.
\label{appendix_target_fidelity_sde}
\end{eqnarray}

For the initial state used in this paper,
$F^c_{{\rm tar}, 0}=1/2$.
More generally, we assume
$0<F^c_{{\rm tar}, 0}<1$.
We introduce the logit variable
\begin{eqnarray}
Y_t
=
\ln
\frac{
F^c_{{\rm tar}, t}
}{1-F^c_{{\rm tar}, t}
}.
\label{appendix_logit}
\end{eqnarray}
$F^c_{{\rm tar}, t}\to1$
is equivalent to
$Y_t\to+\infty$.

Applying It\^o's formula to
Eq.~(\ref{appendix_logit}), we obtain
\begin{eqnarray}
dY_t
=
4\kappa
\left[
\alpha
+
2\tanh
\left(
\frac{Y_t}{2}
\right)
\right]dt
+
4\sqrt{\kappa}dW_t.
\label{appendix_logit_sde}
\end{eqnarray}
Since
$\tanh
\left(Y_t/2\right)
\geq
-1$,
integration of Eq.~(\ref{appendix_logit_sde}) gives
\begin{eqnarray}
Y_t
\geq
Y_0
+
4\kappa
\left(
\alpha-2
\right)t
+
4\sqrt{\kappa}W_t.
\label{appendix_logit_lower_bound}
\end{eqnarray}

For $\alpha>2$, the deterministic term on the right-hand side grows
linearly with time.
Moreover,
$W_t/t\to0$ 
as $t\to\infty$ almost surely.
It follows from
Eq.~(\ref{appendix_logit_lower_bound}) that
$Y_t\to+\infty$
almost surely, and therefore
$F^c_{{\rm tar}, t}\to1$
almost surely.
Thus, the feedback protocol asymptotically stabilizes the target state for
$\alpha>2$.

The target state is not reached at a finite time.
For $0<F_{{\rm tar},0}^{\rm c}<1$, the initial logit $Y_0$ is finite.
Since the drift in Eq.~(B4) is bounded and the diffusion coefficient is
constant, $Y_t$ remains finite at every finite time almost surely.
Therefore, $0<F_{{\rm tar},t}^{\rm c}<1$ at every finite time almost surely,
and the convergence to the target state is asymptotic.

\subsection{Fidelity dynamics and model-specific bounds}
\label{app:qubit_model_derivation}

We write the conditioned state in the Bloch representation as
\begin{eqnarray}
\rho_t^{\rm c}
=
\frac{1}{2}
\left(
I
+
x^c_t\sigma_x
+
y^c_t\sigma_y
+
z^c_t\sigma_z
\right).
\label{appendix_bloch_representation}
\end{eqnarray}
Substituting Eq.~(\ref{appendix_bloch_representation}) and
Eq.~(\ref{qubit_operators}) into the SME gives
\begin{eqnarray}
dx^c_t
&=&
\left(
2u_tz^c_t
-
2\kappa x^c_t
\right)dt
-
2\sqrt{\kappa}\,x^c_tz^c_t\,dW_t,
\label{appendix_bloch_x}
\\
dy^c_t
&=&
-2\kappa y^c_t\,dt
-
2\sqrt{\kappa}\,y^c_tz^c_t\,dW_t,
\label{appendix_bloch_y}
\\
dz^c_t
&=&
-2u_tx^c_t\,dt
+
2\sqrt{\kappa}
\left(
1-(z^c_t)^2
\right)dW_t.
\label{appendix_bloch_z}
\end{eqnarray}

Equation~(\ref{appendix_bloch_y}) is proportional to $y^c_t$.
Since the initial condition satisfies $y^c_0=0$, the trajectory remains in the
$x$-$z$ plane, $y^c_t=0$.
For unit-efficiency monitoring, a pure initial state remains pure under the
conditioned dynamics.
Consequently, $(x^c_t)^2+(z^c_t)^2=1$.

Using
$u_t=-\alpha\kappa x^c_t$,
Eq.~(\ref{appendix_bloch_x}) becomes
\begin{eqnarray}
dx^c_t
=
-2\kappa x^c_t
\left(
1+\alpha z^c_t
\right)dt
-
2\sqrt{\kappa}\,x^c_tz^c_t\,dW_t.
\label{appendix_multiplicative_x}
\end{eqnarray}
The solution can be written as
\begin{eqnarray}
x^c_t
=
x_0
\exp
\Bigg[
-2\kappa
\int_0^t
\left(
1+\alpha z^c_s+(z^c_s)^2
\right)ds
-
2\sqrt{\kappa}
\int_0^t
z^c_s\,dW_s
\Bigg].
\label{appendix_x_solution}
\end{eqnarray}
Since $x_0=1$, Eq.~(\ref{appendix_x_solution}) shows that
$x^c_t>0$
at every finite time almost surely.
The initial state is
\begin{eqnarray}
\rho_0
=
\frac{
I+\sigma_x
}{2}.
\end{eqnarray}
Therefore, the conditioned fidelity with respect to this initial state is
\begin{eqnarray}
F_t^{\rm c}
=
{\rm Tr}
\left(
\rho_0\rho_t^{\rm c}
\right)=
\frac{
1+x^c_t
}{2}.
\label{appendix_initial_fidelity}
\end{eqnarray}
Using
$dF_t^{\rm c}=dx^c_t/2$,
we obtain
\begin{eqnarray}
dF_t^{\rm c}
=
\left(
u_tz^c_t-\kappa x^c_t
\right)dt
-\sqrt{\kappa}\,x^c_tz^c_t\,dW_t.
\end{eqnarray}
After substituting the feedback law, this becomes
\begin{eqnarray}
dF_t^{\rm c}
=
-\kappa x^c_t
\left(
1+\alpha z^c_t
\right)dt
-\sqrt{\kappa}\,x^c_tz^c_t\,dW_t.
\label{appendix_initial_fidelity_sde}
\end{eqnarray}
Hence,
\begin{eqnarray}
A(\rho_t^{\rm c})
&=&
-\kappa x^c_t
\left(
1+\alpha z^c_t
\right),
\\
B(\rho_t^{\rm c})
&=&
-\sqrt{\kappa}\,x^c_tz^c_t.
\end{eqnarray}

The finite-time departure velocity is
\begin{eqnarray}
v_{\rm c}
&=&
-\frac{1}{\tau}
\int_0^\tau
dF_s^{\rm c}
\nonumber\\
&=&
\frac{\kappa}{\tau}
\int_0^\tau
x^c_s
\left(
1+\alpha z^c_s
\right)ds
+
\frac{\sqrt{\kappa}}{\tau}
\int_0^\tau
x^c_sz^c_s\,dW_s.
\end{eqnarray}
This gives the finite-time contributions
$\overline{A}$ and $\overline{B}$ used in the main text.

We next derive model-specific state-independent bounds.
For the diffusion coefficient,
\begin{eqnarray}
B^2(\rho_t^{\rm c})
=
\kappa (x^c_t)^2(z^c_t)^2.
\end{eqnarray}
Using
$(x^c_t)^2+(z^c_t)^2=1$,
we obtain
\begin{eqnarray}
(x^c_t)^2(z^c_t)^2
=
(z^c_t)^2
\left(
1-(z^c_t)^2
\right)
\leq
\frac{1}{4}.
\end{eqnarray}
Therefore,
$B_{\max}^2
=\kappa/4$
is a valid choice.

The instantaneous departure drift is
\begin{eqnarray}
-A(\rho_t^{\rm c})
=
\kappa x^c_t
\left(
1+\alpha z^c_t
\right).
\end{eqnarray}
Since $x^c_t>0$,
$x^c_t=\sqrt{
1-(z^c_t)^2}$.
We therefore define
\begin{eqnarray}
f_\alpha(z)
=
\sqrt{
1-z^2
}
\left(
1+\alpha z
\right),
\end{eqnarray}
where $z\in[-1, 1]$. Its derivative is
\begin{eqnarray}
\frac{df_\alpha}{dz}
=
\frac{
\alpha-z-2\alpha z^2
}{
\sqrt{
1-z^2
}
}.
\end{eqnarray}
The positive stationary point is
\begin{eqnarray}
z_*
=
\frac{
\sqrt{
1+8\alpha^2
}
-1
}{
4\alpha
}.
\label{appendix_zstar}
\end{eqnarray}
For $\alpha>0$, this point gives the global maximum of
$f_\alpha(z)$.
Defining
\begin{eqnarray}
M_\alpha
=
\sqrt{
1-z_*^2
}
\left(
1+\alpha z_*
\right),
\label{appendix_Malpha}
\end{eqnarray}
we obtain
\begin{eqnarray}
-A(\rho_t^{\rm c})
\leq
\kappa M_\alpha.
\end{eqnarray}
Thus, a valid model-specific drift bound is
$A_{\max}=\kappa M_\alpha$.

For comparison, the general spectral estimates derived in
Appendix~\ref{app:state_independent_derivation} give
\begin{eqnarray}
A_{\max}
&=&
\kappa
\left(
\alpha+1
\right), \\
B_{\max}^2
&=&
\kappa.
\end{eqnarray}
The model-specific estimates given above
are tighter because they use the purity of the conditioned state and the
invariant $x$-$z$ plane.


\end{document}